\documentclass[12pt]{article}
\usepackage[utf8]{inputenc}
\usepackage{amsmath}
\numberwithin{equation}{section}
\usepackage{amsfonts}
\usepackage{amssymb}
\usepackage{mathrsfs}
\usepackage{slashed}
\usepackage{indentfirst}
\usepackage{graphics}
\usepackage{hyperref}
\usepackage[framemethod=TikZ]{mdframed}
\mdfdefinestyle{MyFrame}{%
    linecolor=blue,
    outerlinewidth=2pt,
    roundcorner=20pt,
    innertopmargin=\baselineskip,
    innerbottommargin=\baselineskip,
    innerrightmargin=20pt,
    innerleftmargin=20pt,
    backgroundcolor=white!50!white }
      
\begin{document}
\title{Fibre bundle generated theory of higher dimensional gravity}
\author{Theo Verwimp \footnote{Former affiliated with: Physics Department; U.I.A., Universiteit Antwerpen Belgium. On retirement from ENGIE Laborelec, Belgium.}\\
e-mail: \href{mailto:theo.verwimp@telenet.be}{theo.verwimp@telenet.be}}
\renewcommand{\today}{October 14, 2021}
\maketitle
With the help of Weil polynomials on the Lie algebra of the Poincaré group in D dimensions, we construct a Lagrangian form for higher dimensional gravity on a principal fibre bundle whose base space is an even-dimensional Riemann-Cartan spacetime and whose structure group is the Poincaré group.

\section{Introduction}
In a series of papers, an interesting method has been developed for the construction of a Lagrangian form on a principal fibre bundle (PFB) by means of Weil polynomials [1-3]. In a first paper [1], Einstein gravity (with torsion) was reconstructed on a PFB with as structure group the Lorentz group. Subsequently [2], four-dimensional $N=1$ supergravity based on the super Poincaré group, the $OSp(1,4)$ and the $OSp(2,4)$ supergravity theories, were reconstructed using these fibre bundle
methods. If the PFB is the bundle of orthonormal frames over a spacetime of arbitrary even dimension D, which has $SO(1,D-1)$ as structure group, it was shown in ref.[3] how the Lovelock Lagrangian is reconstructed using Weil polynomials on the Lie algebra $so(1,D-1)$.

In the present paper we give a construction parallel to the foregoing, of a Lagrangian form on the PFB over a spacetime of arbitrary even dimension
$D=2m$ with as structure group the Poincaré group. Having regard to the result obtained for the $SO(1,D-1)$ bundle, the present construction gives the most natural generalization of the Lovelock Lagrangian to spacetimes of non-zero torsion. We remark that certain aspects of Lovelock gravity with non-zero torsion have been studied in ref.[4].

\section{Fibre bundle reduction}
Let $M$ be a spacetime of even dimension $D=2m$, and $P(M,G)$ a bundle of affine frames over $M$ with as structural group the Poincaré group
$G=ISO(1,D-1)$. Let $\mathscr{G}$ denote the Lie algebra of $G$. It is defined by
\begin{align}
&\left[J_{ab},J_{cd}\right]=J_{ad}\eta_{bc}+J_{bc}\eta_{ad}-J_{ac}\eta_{bd}-J_{bd}\eta_{ac},\\
&\left[J_{ab},P_{c}\right]=P_{a}\eta_{bc}-P_{b}\eta_{ac},
\end{align} 
and has the direct sum decomposition
\begin{equation}
\mathscr{G}=\mathscr{H}+\mathscr{T}.
\end{equation}
The generators $J_{ab}$ span the subalgebra $\mathscr{H}=so(1,D-1)$ of $\mathscr{G}$ and generate the Lorentz subgroup $H=SO(1,D-1)$ of $G$. The $P_{a}$ span the vector space $\mathscr{T}=\mathbb{R}^{D}_{1,D-1}$. Let $\mu$ be a $\mathscr{G}$-valued connection form in $P$ which has, corresponding to (2.3), the decomposition
\begin{equation}
\mu=\omega+\theta.
\end{equation}
The one-form $\omega$ is a $\mathscr{H}$-valued Lorentz connection and $\theta$ is a one-form on $P$ taking values in $\mathscr{T}$. Explicitly we may write
\begin{equation}
\mu= \frac{1}{2}\omega^{ab}J_{ab}+\theta^{a}P_{a}.
\end{equation}  
The curvature two-form $\Delta$ on $P$ calculated from $\mu$,
\begin{equation}
\Delta=d\mu+\frac{1}{2}\left[\mu,\mu\right],
\end{equation}
with $\left[\;,\;\right]$ denoting the exterior product of forms with values in a Lie algebra, can be written
\begin{equation}
\Delta=\Omega+\Theta=\frac{1}{2}\Omega^{ab}J_{ab}+\Theta^{c}P_{c},
\end{equation}
where
\begin{equation}
\Omega^{ab}=D\omega^{ab}=d\omega^{ab}+\omega^{ac}\wedge\omega_{c}\medskip^{b},
\end{equation}
\begin{equation}
\Theta^{a}=D\theta^{a}=d\theta^{a}+\omega^{ac}\wedge\theta_{c}.
\end{equation}
From the Bianchi identity
\begin{equation}
d\Delta+\left[\mu,\Delta\right]=0
\end{equation}
we find
\begin{equation}
D\Omega^{ab}=d\Omega^{ab}+\omega^{ac}\wedge\Omega_{c}\medskip^{b}-\Omega^{ac}\wedge\omega_{c}\medskip^{b}=0
\end{equation}
\begin{equation}
D\Theta^{a}=d\Theta^{a}+\omega^{a}\medskip_{b}\wedge\Theta^{b}=\Omega^{ab}\wedge\theta_{b}.
\end{equation}

\section{Weil polynomials}
The subgroup $H$ of $G$ acts on $\mathscr{G}$ to the left and forms a linear representation of $H$: $(h,T)\in H\times\mathscr{G}\rightarrow h\cdot T \in\mathscr{G}$ where $h\cdot T=Ad(h)T=hTh^{-1}$ for $T\in\mathscr{H}$ and $h\cdot T=\rho(h)T$ for $T\in\mathscr{T}$, $\rho(h)$ a D-dimensional representation of $H$. We can define a Weil polynomial [5] $L_{m}$ of degree $m$ as a multilinear symmetric real valued function on the Lie algebra $\mathscr{G}$ which is $H$-invariant, i.e.

\begin{equation}
L_{m}(h\cdot T_{1},...,h\cdot T_{m})=L_{m}(T_{1},..., T_{m})
\end{equation}
for all $h\in H$ and $T_{i}\in\left\lbrace T_{a}\right\rbrace=\left\lbrace J_{ab},P_{c}\right\rbrace_{a<b,c=1,...,D}$ a basis for $\mathscr{G}$. The space of Weil polynomials of degree $m$ is denoted by $\smash{S_{H}\medskip^{m}(\mathscr{G})}$.

A Weil polynomial of type (3.1) on $\mathscr{G}$ can always be written as
\begin{equation}
L_{m}(J_{a_{1}b_{1}}, ...,J_{a_{k}b_{k}}, P_{a_{k+1}}, ...,P_{a_{m}}), \qquad k\in \left\lbrace 0,1, ...,m\right\rbrace.
\end{equation}
The Weil polynomials $L_{m}(J_{a_{1}b_{1}}, ...,J_{a_{m}b_{m}})$ (corresponding to $k=m$ in (3.2)) are polynomials on the subalgebra $\mathscr{H}$ of $\mathscr{G}$. They were calculated in ref.[3]:
\begin{equation}
L_{m}(J_{a_{1}a_{2}}, ...,J_{a_{D-1}a_{D}})=2^{m}(C_{1}\epsilon_{a_{1}...a_{D}}+C_{2}\eta_{a_{1}...a_{m},a_{m+1}...a_{D}}),
\end{equation}
where 
\begin{align*}
&C_{1}=(m!2^{m})^{-1}\\
&C_{2}=\left[(m/2)!2^{m/2}\right]^{-2}, \qquad \textrm{if m is even},\\
&\quad\; =0, \qquad\qquad\qquad\qquad\; \textrm{if m is uneven},\quad\quad\quad\quad\quad\quad\quad\\
&\epsilon_{a_{1}...a_{D}}\;\textrm{is the totally antisymmetric tensor with}\;\epsilon_{1...D}=1, 
\end{align*}
\begin{equation}
\eta_{a_{1}...a_{k},b_{1}...b_{k}}=\sum_{\sigma\in S_{k}}\epsilon(\sigma)\eta_{a_{1}b_{\sigma(1)}}... \eta_{a_{k}b_{\sigma(k)}},\quad\quad\quad\quad\quad\quad\quad\quad
\end{equation}
where the summation is taken over all permutations $\sigma$ of $(1, ...,k)$ and $\epsilon(\sigma)$ is the sign of the permutation.

Next, we show that for $k=0$ in (3.2)
\begin{equation}
\begin{aligned}
&L_{m}(P_{a_{1}}, ..., P_{a_{m}})=C_{3}\eta_{(a_{1}a_{2}}\eta_{a_{3}a_{4}}\cdot\cdot\cdot\eta_{a_{m-1}a_{m})}, \quad \textrm{if m is even},\\
&\quad\quad\quad\quad\quad\quad\quad =0,\quad\quad \quad\quad\quad\quad\quad\quad\quad\quad\quad\;\textrm{if m is uneven.}
\end{aligned}
\end{equation}
Therefore, let $h=\textrm{exp}(\sum_{a<b}t^{ab}J_{ab})$ and apply $d/dt^{ab}\mid_{t=0}$ to
\begin{equation}
L_{m}(h\cdot P_{a_{1}},h\cdot P_{a_{2}}, ..., h\cdot P_{a_{m}})=L_{m}(P_{a_{1}}, ..., P_{a_{m}}).
\end{equation}
We find
\begin{equation}
\sum_{i=1}^{m}(J_{ab})^{b_{i}}\medskip_{a_{i}}L_{a_{1}\cdot\cdot\cdot b_{i}\cdot\cdot\cdot a_{m}}=0\quad (b_{i}\;\textrm{on the \textit{i}th place}), 
\end{equation}
Where $L_{a_{1}\cdot\cdot\cdot a_{m}}\equiv L_{m}(P_{a_{1}}, ..., P_{a_{m}})$ and $(J_{ab})^{b_{i}}\medskip_{a_{i}}$ are the generators of $\mathscr{H}=so(1,D-1)$ in $D\times D$-matrix representation:
\begin{equation}
(J_{ab})^{b_{i}}\medskip_{a_{i}}=\delta^{b_{i}}\medskip_{a}\eta_{ba_{i}}-\delta^{b_{i}}\medskip_{b}\eta_{aa_{i}}.
\end{equation}
Substitution of eq.(3.8) into eq.(3.7) yields after multiplication by $\eta^{ba_{1}}$
\begin{align}
&L_{a_{1}}=0,\;\;\textrm{if}\;m=1,\\
&L_{a_{1}a_{2}}\propto\eta_{a_{1}a_{2}}=\eta_{(a_{1}a_{2})} \;\;\textrm{if}\;m=2\quad\textrm{(see also ref.[1])},\\
&L_{a_{1}\cdot\cdot\cdot a_{m}}\propto\sum_{i=2}^{m}\eta_{a_{1}a_{i}}L_{a_{2}\cdot\cdot\cdot a_{i-1}a_{i+1}\cdot\cdot\cdot a_{m}},\;\;\textrm{if}\;m\geq 3.
\end{align}
The result (3.5) then follows by induction.

If $k\neq 0, m$ in (3.2)  it is clear that no Lorentz invariant can be constructed (from Minkowski and antisymmetric tensor) that meets the required symmetry properties, except for the zero tensor. For the case when $m=2$ this was explicitly shown in ref.[1].

\section{A Lagrangian form on the Poincaré bundle}
If $\psi_{1},...,\psi_{m}$ are $\mathscr{G}$-valued two-forms on $P$, then we define for each $L_{m}\in S^{m}_{H}(\mathscr{G}) $ a real valued Weil form $L_{m}(\psi_{1},...,\psi_{m})$ on $P$ of degree $2m=D$ the dimension of $M$, by
\begin{multline}
L_{m}(\psi_{1},...,\psi_{m})(X_{1},...,X_{D})=\\\left( \frac{1}{2}\right)^{m}\sum_{\sigma\in S_{D}}\epsilon(\sigma)L_{m}\left(\psi_{1}(X_{\sigma(1)},X_{\sigma(2)}),...,\psi_{m}(X_{\sigma(D-1)},X_{\sigma(D)})\right) 
\end{multline}
for $X_{1},...,X_{D}\in T_{u}(P)$ (tangent space of $P$ at $u$), where the summation is taken over all permutations $\sigma$ of $1,...,D)$ and $\epsilon (\sigma)$ is the sign of the permutation. If $\left\lbrace T_{A}\right\rbrace $ is a basis for $\mathscr{G}$ such that $\psi_{i}=\psi_{i}^{A}T_{A}$ [$\psi_{i}^{A}\in\Lambda^{2}(P,\mathbb{R})$ the space of real-valued two-forms on $P$], then one obtains from the multilinearity of $L_{m}$ and the definition of the wedge product that
\begin{equation}
L_{m}(\psi_{1},...,\psi_{m})=L_{m}(T_{A_{1}},...,T_{A_{m}})\psi_{1}^{A_{1}}\wedge\cdot\cdot\cdot\wedge\psi_{m}^{A_{m}}.
\end{equation}

In analogy with ref.[3] we define on $P$ a Lagrangian form (a $D=2m$-form) of the type
\begin{equation}
L_{m}(\Delta +\bar{\alpha}f(\theta))=L_{m}(\Delta +\alpha_{1}f(\theta), ...,\Delta +\alpha_{m}f(\theta))
\end{equation}
where

(i) $L_{m}$  is the Weil polynomial formed by summation over all algebraically independent elements of $S^{m}_{H}(\mathscr{G})$,

(ii) $\Delta$ is the curvature two-form given by eq.(2.7),

(iii) $f(\theta)=\frac{1}{2}\theta^{a}\wedge\theta^{b}J_{ab}$,

(iv) $\alpha_{j},\, j=1, ...,m$ are constants of dimension (length)$^{-2}$.

From the multilinearity and symmetry of the Weil form (4.3) we obtain after substitution of eq.(2.7) and making use of eq.(4.2),
\begin{multline}
L_{m}(\Delta +\bar{\alpha}f(\theta))=\sum_{p=0}^{m}\tilde{\alpha}_{p}\sum_{q=0}^{p}\beta_{q}L_{m}(J_{a_{1}a_{2}}, ...,J_{a_{2q-1}a_{2q}}, P_{b_{1}}, ..., P_{b_{p-q}}, J_{c_{1}c_{2}}, ..., \\
J_{c_{2(m-p)-1}c_{2(m-p)}})\Omega^{a_{1}a_{2}}\wedge\cdot\cdot\cdot\wedge\Omega^{a_{2q-1}a_{2q}}\wedge\Theta^{b_{1}}\wedge\cdot\cdot\cdot\wedge\Theta^{b_{p-q}}\wedge\theta^{c_{1}}\wedge \cdot\cdot\cdot\wedge\theta^{c_{2(m-p)}},  
\end{multline}
where
\begin{align}
&\tilde{\alpha}_{p}=\sum_{i_{1}<\cdot\cdot\cdot<i_{m-p}}\alpha_{i_{1}}\times\cdot\cdot\cdot\times\alpha_{i_{m-p}}\quad (p<m);\quad \tilde{\alpha}_{m}=1,\\
&\beta_{q}=2^{p-q-m}\left( \begin{matrix}
 p \\ 
 p-q
 \end{matrix}\right).
\end{align}
From the results of section 3, we have that in eq.(4.4) only the terms with $p=q$ and the term with $p=m,\; q=0$ will contribute. If we also use
$\theta^{a}\wedge\theta^{b}\eta_{ab}=0$, we obtain the final expression for the Lagrangian form on $P$:
\begin{multline}
\mathscr{L}=\left(\sum_{p=0}^{m}C_{1}\tilde{\alpha}_{p}\epsilon_{a_{1}\cdot\cdot\cdot a_{D}}+\sum_{p=m/2}^{m}C_{2}\tilde{\alpha}_{p}\eta_{a_{1} ...  a_{m},a_{m+1}...a_{D}}\right)\Omega^{a_{1}a_{2}}\wedge\cdot\cdot\cdot\wedge\Omega^{a_{2p-1}a_{2p}}\\ \wedge\theta^{a_{2p+1}}\wedge\cdot\cdot\cdot\wedge\theta^{a_{D}}+C_{3}\Theta^{a_{1}}\wedge\cdot\cdot\cdot\wedge\Theta^{a_{m/2}}\wedge\Theta_{a_{1}}\wedge\cdot\cdot\cdot\wedge\Theta_{a_{m/2}}.
\end{multline}

\section{Reduction of the Poincaré bundle}
To interpret the obtained Lagrangian form $\mathscr{L}$ on $P$  as a gravitational Lagrangian, we have to identify the sub-bundle $P'(M,H)$ of $P$ as the bundle $F(M)$ of orthonormal frames over $M$ and $\mu$ as the connection form of an affine connection $\tilde\Gamma$ on $M$. In that
case we have the homomorphism (see ref.[5], ch. III)
\begin{equation}
\gamma:F(M)\rightarrow P(M,G):(X_{1}, ..., X_{D})\shortmid\rightarrow (O_{x};X_{1}, ..., X_{D}),
\end{equation}
where $(X_{1}, ..., X_{D})$ is a linear frame of $M$ and $O_{x}$ is the origin of $T_{x}(M)$, such that the pullback $\gamma^{\ast}\mu=\mu\vert_{F(M)}$ decomposes as in eq.(2.4), i.e.
\begin{equation}
\mu\vert_{F(M)}=\omega+\theta,
\end{equation}
with $\omega$ a connection form defining a connection $\tilde{\Gamma}$ in
the bundle $F(M)$ and $\theta$ the canonical form of $F(M)$. If correspondingly the curvature tensor $\Delta$ in eq.(2.7) is restricted to $F(M)$, we can identify $\Theta$ and $\Omega$ as the torsion form and the curvature form of the linear connection $\Gamma$. With this restriction understood, the Lagrangian form (4.7) reduces to a gravitational Lagrangian on the orthonormal frame bundle and will
project to a unique D-form $\tilde{\mathscr{L}}$ on $M$ such that
\begin{equation}
\mathscr{L}=\pi^{\ast}\tilde{\mathscr{L}},
\end{equation}
with $\pi:F(M)\rightarrow M$ the projection on base manifold $M$ (see refs. [1,3]).

The reduction described above works not only for the Poincaré group, but also for the de Sitter group of type either $(1,D)$ or $(2,D-1 )$[6]. In fact, with $SO(2,D-1)$ in the role of $G$, the Lie algebra $\mathscr{G}$
defined by the commutation relations (2.1) and (2.2) supplemented with
\begin{equation}
\left[P_{a},P_{b}\right]=\alpha J_{ab}
\end{equation}
where $\alpha^{-1/2}$ is the de Sitter length. The curvature two-form $\tilde{\Delta}$ of the anti-de Sitter bundle is now
\begin{equation}
\tilde{\Delta}=\Omega+\Theta+\frac{1}{2}\left[\theta,\theta\right],
\end{equation}
such that the Weil form $L_{m}(\tilde{\Delta}, ...,\tilde{\Delta})$ now produces the Lagrangian form given in eq.(4.7). The coupling constants $\tilde{\alpha}_{p}$ are then given by
\begin{equation}
\tilde{\alpha}_{p}=\left( \begin{matrix}  m \\  m-p \end{matrix}\right)\alpha^{m-p},
\end{equation}
i.e., $\alpha=\alpha_{1}=...=\alpha_{m}$ in eq.(4.5).

\section{Discussion}
The present study found its motivation in the search for a more natural generalisation of Einstein-Cartan theory to higher dimensional gravity than the theory obtained by dimensional continuation of the Einstein-Cartan Lagrangian. This last theory was used for example in refs.[7] and [8], in the study of compactifying solutions of the form $M^{4}\times B^{n}$, $B^{n}$ a compact $n$-dimensional space of non-zero torsion.
For the compactification and dimensional reduction, within the higher dimensional theory given here, see ref.[9] for the case of zero torsion.

Lagrangian forms on a PFB, such as (4.7), are invariant under local symmetry transformations. In fact the invariance of (super)-Poincaré transformations has been discussed in detail by Yates and Awada [10] (see also ref.[2]). Neglecting supersymmetry considerations, the proof of the invariance of (4.7) under local Poincaré transformations is a straightforward extension of the result in ref.[10].

The construction of the invariant gravitational Lagrangian form on the Poincaré bundle is essentially a generalisation to spacetimes with arbitrary torsion of the methods developed in ref.[3] concerning the
construction of an invariant Lagrangian form on the orthonormal frame bundle. But in view of the remark at the end of section 5, it also extends to arbitrary even dimensions, the MacDowell-Mansouri (MM) construction of Einstein gravity (a comprehensive discussion of which can be found in ch.21 of ref.[11]). Contrary to the MM-construction, we
do not exclude parity violating terms a priori. The result is that only in $D=4k$ dimensions (k=1,2,...) we obtain a parity violating gravity, since in that case $C_{2}$ and $C_{3}$ are non-zero. Here, solutions without definite space inversion properties for the internal dimensions are possible. If $D=4k+2$, the Lagrangian (4.7) is invariant under space inversions and reduces then to the "Lovelock-Cartan" Lagrangian
supplemented with the Euler topological invariant (i.e., the term with $p=m$ in (4.7)). We also remark that the methods given here, together with those of ref.[2], set the stage for the construction of the
higher dimensional supergravity theory based on the supersymmetric extension of the dimensionally extended Euler densities.

By construction, our invariant Lagrangian (4.7) is a polynomial in $\Omega$, $\Theta$, $\theta$, i.e.,
\begin{equation}
\mathscr{L}=\mathscr{L}(\omega,d\omega,\theta,d\theta).
\end{equation}
However, the Lorentz connection $\omega$ splits as [12]
\begin{equation}
 \omega^{ab}=\tilde{\omega}^{ab}+\tau^{ab},
\end{equation} 
where $\tilde{\omega}^{ab}$ is the unique Levi-Cevita connection and $\tau^{ab}$
 the contorsion term. Therefore after translating to second-order formalism, one finds that
\begin{equation}
\mathscr{L}=\mathscr{L}(g,\partial g,\partial\partial g,T,\partial T),
\end{equation}
where $g$ is the metric tensor on $M$, compatible with the connection $\omega$ and $T$ is the torsion tensor.

In general, the field equations are polynomials in $\Omega$, $\Theta$, $D\Theta$ and $\theta$. Call the first, second and third term in (4.7) $\mathscr{L}_{1}$,  $\mathscr{L}_{2}$,  $\mathscr{L}_{3}$ respectively. Variation of $\mathscr{L}_{1}$ with respect to $\theta$ and $\omega$ gives respectively[13] 

\begin{multline}
\delta_{\theta}\mathscr{L}_{1}=\sum_{p=0}^{m-1}C_{1}(D-2p)\tilde{\alpha}_{p}\delta\theta^{a_{2p+1}}\wedge\epsilon_{a_{1}...a_{D}}\Omega^{a_{1}a_{2}}\wedge\cdot\cdot\cdot\wedge\Omega^{a_{2p-1}a_{2p}}\\ \wedge\theta^{a_{2p+2}}\wedge\cdot\cdot\cdot\wedge\theta^{a_{D}}
\end{multline}
\begin{multline}
\,\delta_{\omega}\mathscr{L}_{1}=\sum_{p=1}^{m-1}C_{1}p(D-2p)\tilde{\alpha}_{p}\delta\omega^{a_{1}a_{2}}\wedge\epsilon_{a_{1}...a_{D}}\Omega^{a_{3}a_{4}}\wedge\cdot\cdot\cdot\wedge\Omega^{a_{2p-1}a_{2p}}\\ \wedge\Theta^{a_{2p+1}}\wedge\theta^{a_{2p+2}}\wedge\cdot\cdot\cdot\wedge\theta^{a_{D}} + \textrm{exact form}.
\end{multline}
Varying $\mathscr{L}_{2}$ gives similar polynomials, now involving the $\eta_{a_{1} ...  a_{m},a_{m+1}...a_{D}}$ tensor. Finally,
\begin{align}
&\delta_{\theta}\mathscr{L}_{3}=mC_{3}\delta\theta^{a_{1}}\wedge D(\Theta^{a_{2}}\wedge\cdot\cdot\cdot\wedge\Theta^{a_{m/2}}\wedge\Theta_{a_{1}}\wedge\cdot\cdot\cdot\wedge\Theta_{a_{m/2}})+\textrm{exact form}\\
&\delta_{\omega}\mathscr{L}_{3}=mC_{3}\delta\omega^{a_{1}b}\wedge\theta_{b}\wedge \Theta^{a_{2}}\wedge\cdot\cdot\cdot\wedge\Theta^{a_{m/2}}\wedge\Theta_{a_{1}}\wedge\cdot\cdot\cdot\wedge\Theta_{a_{m/2}}
\end{align}
However, by the Bianchi identity (2.12), these are polynomials in $\Omega$, $\Theta$ and $\theta$ such that the field equations are second order in $g$ and first order in $T$. From the foregoing we see that $\omega$ is determined dynamically by the field equations, unless $D=4$. In that case, the second field equation is algebraic in $\omega$:
\begin{equation}
\delta_{\omega}\mathscr{L}=\delta_{\omega}\mathscr{L}(\omega, \theta, d\theta).
\end{equation}
In second order formalism, this means algebraic in torsion. The consequence of this fact, non-propagating torsion in a four-dimensional spacetime, has been emphasized by Hehl et al. [14].

\newpage 
\begin{center}
\textbf{\Large References}
\end{center}
\hspace*{2mm}$[1]$ K. Kakazu and S. Matsumoto, Prog. Theor. Phys. 78 (1987) 166.\\
\hspace*{2mm}$[2]$ K. Kakazu and S. Matsumoto, Prog. Theor. Phys. 78 (1987) 932; 79 (1988) 1431.\\
\hspace*{2mm}$[3]$ T. Verwimp, Prog. Theor. Phys. 80 (1988) 330; \href{https://arxiv.org/abs/2106.07508}{arXiv:2106.07508 [gr-qc]}.\\
\hspace*{2mm}$[4]$ J.T. Wheeler, Nucl. Phys. B 273 (1986) 732; D.H. Tchrakian, Class. Quantum 

\hspace*{1,5mm}Grav. 4 (1987) L217.\\
\hspace*{2mm}$[5]$ S. Kobayashi and K. Nomizu, Foundations of differential
geometry I (Interscience, 

\hspace*{1,5mm}New York, 1963).\\
\hspace*{2mm}$[6]$ P.K. Smrz, J. Math. Phys. 28 (1987) 2824.\\
\hspace*{2mm}$[7]$ B. Mclnnes, ICTP Trieste, preprint IC/85/58.\\
\hspace*{2mm}$[8]$ N.A. Batakis, K. Farakos, D. Kapetanakis, G. Koutsoumbas and G. Zoupanos, Phys. 

\hspace*{1,5mm}Lett. B220 (1989) 513.\\
\hspace*{2mm}$[9]$ T. Verwimp, Class. Quantum Grav. 6 (1989) 1655-1663;  \href{https://arxiv.org/abs/2108.10633}{arXiv:2108.10633 [gr-qc]}.\\
$[10]$ R.G. Yates, Commun. Math. Phys. 76 (1980) 255; M.A. Awada, Commun. Math. 

\hspace*{1,5mm}Phys. 91 (1983) 53.\\
$[11]$ P.G.O. Freund, Introduction to supersymmetry (Cambridge
Univ. Press, Cambridge,

\hspace*{1,5mm}1986).\\
$[12]$ R.P. Wallner, Gen. Rel. Grav. 12 (1980) 719.\\
$[13]$ F. Müller-Hoissen, Phys. Lett. B 163 (1985) 106.\\
$[14]$ F.W. Hehl, P. von der Heyde and G.D. Kerlick, Rev. Mod. Phys. 48 (1976) 393.

\end{document}